\documentclass[%
 reprint,
 onecolumn,
 amsmath,amssymb,
 aps,
floatfix,
]{revtex4-2}

\usepackage{graphicx}% Include figure files
\usepackage{dcolumn}% Align table columns on decimal point
\usepackage{bm}% bold math
\usepackage{float}
\usepackage[dvipsnames]{xcolor}
\usepackage{subfig}

\begin{document}

\preprint{APS/123-QED}

\title{Mitigating Disinformation in Social Networks through Noise}% Force line breaks with \\
%\thanks{A footnote to the article title}%

\author{Diana Riazi}
 %\altaffiliation{Department of Computer Science, University College London.}%Lines break automatically or can be forced with \\
%\author{}%
 %\email{Second.Author@institution.edu}
\affiliation{Department of Computer Science, University College London, 66-72 Gower Street, WC1E 6EA London, United Kingdom}

\author{Giacomo Livan}
\affiliation{Dipartimento di Fisica, Universit\`a di Pavia, Via A. Bassi 6, 27100 Pavia, Italy}
\affiliation{Department of Computer Science, University College London, 66-72 Gower Street, WC1E 6EA London, United Kingdom}

%\collaboration{MUSO Collaboration}%\noaffiliation

%\author{Charlie Author}
% \homepage{http://www.Second.institution.edu/~Charlie.Author}
%\affiliation{
% Second institution and/or address\\
% This line break forced% with \\
%}%
%\affiliation{
% Third institution, the second for Charlie Author
%}%
%\author{Delta Author}
%\affiliation{%
% Authors' institution and/or address\\
% This line break forced with \textbackslash\textbackslash
%}%

%\collaboration{CLEO Collaboration}%\noaffiliation

\date{\today}% It is always \today, today,
             %  but any date may be explicitly specified

\begin{abstract}
An abundance of literature has shown that the injection of noise into complex socio-economic systems can improve their resilience. This study aims to understand whether the same applies in the context of information diffusion in social networks. Specifically, we aim to understand whether the injection of noise in a social network of agents seeking to uncover a ground truth among a set of competing hypotheses can build resilience against disinformation. We implement two different stylized policies to inject noise in a social network, i.e., via random bots and via randomized recommendations, and find both to improve the population's overall belief in the ground truth. Notably, we find noise to be as effective as debunking when disinformation is particularly strong. On the other hand, such beneficial effects may lead to a misalignment between the agents' privately held and publicly stated beliefs, a phenomenon which is reminiscent of cognitive dissonance.

\end{abstract}

%\keywords{Suggested keywords}%Use showkeys class option if keyword
                              %display desired
\maketitle

%\tableofcontents

\section{\label{sec:level1}Introduction}

Across a variety of literatures and contexts, the role of noise and randomness have been shown to have beneficial effects in complex systems~\cite{mantegna1996noise,albert2000error,biondo2013beneficial,livan2019don}, i.e., systems made of large numbers of interacting components giving rise to non-trivial emergent macroscopic properties.

In fact, complex systems often experience a tension between efficiency and resilience, with the former often being a desired property of the system, which, however, can lead build up its fragility and ultimately lead to unintended consequences (see, e.g.,~\cite{bardoscia2017statistical}). A prime example of such tension that was recently on display is supply chains, which underwent chaos during the Covid-19 pandemic due to their lack of robustness to shocks, induced by a decades-long push towards increasingly high efficiency levels~\cite{de2022examining}. Similarly, recent work has put forward evidence in support of the `Peter Principle'~\cite{pluchino2010peter}, i.e., that random promotion strategies have the potential to prevent the diffusion of incompetence in hierarchical organizations. In line with this result, the efficiency of Parliament has been shown to improve by the apparent existence of an optimal random selection of legislators within a two-party system, as initially explored in \cite{pluchino2011accidental}. Moreover, in the context of financial markets, standard and popular strategies, on average, do not perform better compared to random trading strategies~\cite{biondo2013beneficial}, which actually have a stabilizing effect on market dynamics~\cite{farmer2005predictive}. 

Given the benefits that the injection of noise has towards building the resilience of both physical and socio-economic complex systems, in this paper we seek to adopt the same logic of many of the studies mentioned above, in order to determine whether `the right amount' of noise can lead to beneficial effects against the diffusion of disinformation in online social networks (OSNs), a well-known plague of our time~\cite{vosoughi2018spread}. In fact, on the one hand OSNs are extremely efficient at aggregating the sparse information available to their users. At the same time, we now know very well how such efficiency can backfire in a variety of ways~\cite{zollo2017debunking}, or even be leveraged by malicious agents to spread `fake news' and polarize users~\cite{sikder2020minimalistic}.

In the following, we will investigate with numerical simulations whether the presence of noisy agents --- or, as it is known in the distributed systems literature, \textit{fault-tolerant agents}~\cite{mitra2020new,su2019defending} --- may contribute to mitigating the spread of misleading information, and how the injection of noise would compare against more conventional --- yet not necessarily effective --- ways of combating the spread of misleading information (e.g. fact-checking~\cite{li2023combating} or debunking~\cite{chan2017debunking}).

To explore this question, we adopt the the social learning framework introduced in~\cite{lalitha2018social} --- known as distributed hypothesis testing (DHT) --- and expanded upon in~\cite{riazi2024public}, which will allow us to define simple measures of truthfulness (i.e., a society's ability to find consensus on a ground truth) and cognitive dissonance (i.e., the mental toll experienced by agents exposed to contrasting information). In the following, we provide a brief overview of this framework.

\section{\label{sec:level3}Social Learning and Distributed Hypothesis Testing}

The aim in DHT models is for agents to learn a ground truth, amidst a collection of competing hypotheses or narratives, through the aggregation over time of privately signals. These are received as random draws from a distribution governed by the ground truth. However, agents are unaware of this, and the learning process intends at recognizing from which distribution they are receiving signals among an array of different distributions, encoding the different competing hypotheses.  Distributions may be interpreted as the uncertainty which surrounds such hypotheses or perspectives. In the following, we provide a brief overview of DHT, as introduced in~\cite{lalitha2018social}, and recall the definitions of truthfulness and cognitive dissonance, that we will employ in following, as introduced in~\cite{riazi2024public}. 

We consider $N$ agents connected by a network $W$ of interactions (such that $W_{ij} > 0$ when agents $i$ and $j$ are connected and $W_{ij} = 0$ otherwise). The weight $W_{ij}$ represents the amount of influence that agent $i$ accepts from agent $j$. Because of this, the matrix $W$ is assumed to be row-stochastic, i.e. $\sum_{j=1} ^{N} W_{ij}=1$. Let us also consider a set of $N \times M$ multivariate distributions $f_i(X; \theta_{ik})$ ($i = 1, \ldots, N$, $k = 1, \ldots, M$), where $\theta_{ik}$ denotes the set of parameters that define the $k$-th distribution associated with agent $i$. To simplify notation, in the following we will drop the subscript $i$, as it will be clear from context to which agent the set of parameters $\theta_{ik}$ refers to.

The $M$-th distribution is the one corresponding to the ground truth for each agent. Time is discrete and denoted as $t=1,...,T$. At time $t = 0$ each agent is initialized with a random vector of private beliefs $\boldsymbol{q}_i^{(0)} = \left (q_i^{(0)}(\theta_1), \ldots, q_i^{(0)}(\theta_M) \right )$, such that $q_i^{(0)}(\theta_k) \geq 0, \ \forall k$ and $\sum_{k=1}^M q_i^{(0)}(\theta_k) = 1$. At each time step, we perform the following steps: 

\begin{itemize}
    \item Each agent $i$ ($i = 1, \ldots, N$) receives a signal 
    $X_i^{(t)}$ as a random draw from the distribution corresponding to the ground truth, i.e., $X_i^{(t)}\sim f_i (\cdot ; \theta_M )$.
    \item Each agent $i$ performs a local Bayesian update on their current vector of private beliefs $\boldsymbol{q}_i^{(t)}$ to form a public belief vector $\boldsymbol{b}_i^{(t)}$ with components 
    \begin{equation} \label{eq:pub_beliefs}
        b_{i}^{(t)}(\theta_k) = \frac{f_{i}(X^{(t)}_{i}; \theta_k) \ q_i^{(t-1)}(\theta_k)}{\sum_{\ell = 1}^M f_i(X^{(t)}_{i};\theta_\ell) \ q^{(t-1)}_{i}(\theta_\ell)}
    \end{equation}            
    \item Each agent $i$ shares their public belief vector $\boldsymbol{b}_i^{(t)}$ with all their neighbors in the network, and similarly receives public belief vectors from each of them.
    \item Each agent $i$ updates their private belief vector $\boldsymbol{q}_i^{(t)}$ by averaging the log-beliefs they received from neighbors, i.e.,
    \begin{equation} \label{eq:pvt_beliefs}
q^{(t)}_{i}(\theta_k)=\frac{\exp \left (\sum_{j=1}^N W_{ij} \log b^{(t)}_{j}(\theta_k) \right)}{\sum_{\ell = 1}^M \exp \left (\sum_{j=1}^N W_{ij}\log b^{(t)}_{j}(\theta_\ell) \right )} \ .
    \end{equation}
\end{itemize}

Note that both Eqs.~\eqref{eq:pub_beliefs} and~\eqref{eq:pvt_beliefs} ensure that the public and private belief vectors of each agent remain correctly normalized as probability vectors at each time step.

As shown in~\cite{lalitha2018social}, the ground truth is collectively learnt by all agents exponentially fast under global distinguishability, which characterizes the concept that at least one agent in the network is able to differentiate between any pair of competing hypotheses, i.e., for all $k \neq \ell$, there exists at least one agent $i$ such that $D_\mathrm{KL}(f_i(\cdot,\theta_k) || f_i(\cdot,\theta_\ell)) > 0$, where $D_\mathrm{KL}(\cdot || \cdot)$ denotes the Kullback-Leibler divergence.
%%%%%%%%%%%%%%%%%

\section{\label{sec:level4}Measures of Truthfulness and Cognitive Dissonance}

To characterize the overall network's ability to learn the ground truth, we make use of the following metric introduced in~\cite{riazi2024public}:
 
\begin{equation} \label{eq:truthfulnes}
    \tau(t) =\frac{1}{N}\sum_{i=1}^N q_i^{(t)}(\theta_M) \ ,
\end{equation}
which quantifies the average private belief placed by the agents on the ground truth (recall that by convention the assumption of the $M$-th hypothesis to be the true one). The quantity in Eq.~\eqref{eq:truthfulnes} is referred as \emph{truthfulness}. Truthfulness can also be expressed as the difference between one and the average private belief collectively placed on the $M-1$ wrong hypotheses due to the normalization of private vectors:
\begin{equation} \label{eq:CD}
    \tau (t)= 1 - \frac{1}{N}\sum_{i=1}^N \sum_{\ell=1}^{M-1}
    q_i^{(t)}(\theta_\ell) \ .
\end{equation}

The concept of  cognitive dissonance (CD) in Psychology refers to the mental toll experienced by an individual when faced with contradictory information~\cite{cooper2019cognitive}. In the context of our model, we equate CD to the difference (in absolute value) between what an agent privately believes and what they publicly express. Mathematically, the CD experienced by an agent $i$ at time $t$ regarding some hypothesis $\theta_k$ is
\begin{equation}
C^{(t)}_i(\theta_{k})= \left |q^{(t)}_i(\theta_k)-b^{(t)}_i(\theta_k) \right | \ .
\end{equation}

We recall that the conspirators sustain their
public belief to be one of the $M-1$ wrong hypotheses artificially high, thus effectively spreading disinformation and affecting the `regular' agents' learning process. Specifically, it is imposed that conspirator agents maintain a public belief vector $\boldsymbol{b}^{(t)} = (b_1,b_2,b_3,\ldots)$ with $b_1 \approx 1$ and $b_j \ll 1$ for $j = 2,\ldots, M$, that is, conspirator agents push one of the wrong hypotheses (which, without loss of generality, we assume to be the first one) at each time step.

In the next section, we seek to investigate the role noise plays on such notions, and whether there exists any beneficial effects as shown in other contexts. In other words, does the invocation of noise alleviate disinformation spread?

\section{\label{sec:level5}Policies}
We consider two policies or iterations of noise. Firstly, we consider introducing noise by injecting a fraction $\beta_r$ of agents whose beliefs are determined by random draws. More specifically, we assign those agents' public beliefs (see~\eqref{eq:pub_beliefs}) to be random probability vectors, i.e., vectors with non-negative entries that sum up to one. This is achieved by sampling from the Dirichlet distribution \cite{gouda2004new}, which we will use in the following to refer to this policy. Such a scenario can be likened to a context of individuals interacting with \textit{social bots} in online social network platforms. For instance, some efforts of this literature aim to gauge the susceptibility \cite{wagner2012social} of agents in the presence of attacks which aim to influence, potentially given a malicious agenda. It is worth mentioning that there is much effort on investigating the detection of social bots \cite{ferrara2016rise, keller2019social}, where much is focused naturally on such bots with malicious intent. It should be noted the nuance of the scenario being considered here, namely the distinction between what we call conspirator agents (purposely pumping in a non-truth) and noisy agents (agents whose beliefs are determined `randomly'), denoted respectively by $\beta_c$ and $\beta_r$.  

We also consider simulating noise through topological means, that is through the links in the network being \textit{rewired}. More specifically, following a procedure similar to the ones used in~\cite{livan2017excess,li2019reciprocity}, at each time step we randomly select two pairs of nodes $(i,j)$ and $(k,\ell)$ with $i \neq j \neq k \neq \ell$ such that $W_{ij} > 0$, $W_{k\ell} > 0$, $W_{i\ell} = W_{kj} = 0$, delete the existing links (i.e., we set $W_{ij} = W_{k\ell} = 0$) and place their weights on $W_{i\ell}$ and $W_{kj}$, respectively. Such rewiring protocol preserves the degree of each node, i.e., their number of neighbors in the network, but obviously changes who those neighbors are. Such a form of noise may be seen to parallel social media platform algorithms that recommend content which may be outside what an individual typically sees or with what or whom they interact. For instance, Instagram recently came under criticism for the sentiment of users' feed boasting recommendations over the content of whom they follow \footnote{“Instagram Pauses Updates Following Criticism about Being Too Video Focused.” The Independent, Independent Digital News and Media, 29 July 2022, www.independent.co.uk/tech/instagram-video-update-tiktok-criticism-b2133731.html.}. In essence, this rewiring policy may be considered in conjunction to the literature of recommender systems. More specifically, recommender systems have been criticized for generating a space that allows for filter bubbles, weakening diversity~\cite{kunaver2017diversity}, and potentially giving rise to  confirmation bias~\cite{bhadani2021biases}. Additionally, it has also been shown how misinformation may propagate via recommender systems (e.g. through feedback loop mechanisms~\cite{fernandez2021analysing}). Given such dynamics in the recommender systems' literature, it is favorable to make such topological considerations as carried out in this paper. 

In the following, we will compare the above two policies in their effectiveness at improving truthfulness and/or reducing cognitive dissonance in a variety of networks and scenarios. In order to provide a fair comparison, we will perform comparisons when the two policies inject similar levels of noise in the system. In order to capture that, we will use the same parameter $\beta_r$ to quantify the amount of noise being injected. In the context of the Dirichlet policy, $\beta_r$ will quantify the fraction of noisy agents in the network; in the context of the rewiring policy, instead, $\beta_r$ will quantify the fraction of links being rewired at each time step in our simulations.

\section{\label{sec:level5}Results}
We observe and interpret the effects that the aforementioned policies have on our defined concepts of truthfulness~\eqref{eq:truthfulnes} and cognitive dissonance~\eqref{eq:CD}. 

For the sake of simplicity, we first investigate such effects within Erdos-Rényi (ER) networks, i.e., networks where each pair of nodes has the same probability of being connected, which --- in the limit of large networks --- gives rise to a Poissonian degree distribution. Though ER networks do not reflect the heterogeneity of the degree distributions of real-world social networks (which we will address later), they provide of course the advantage of straightforwardly understanding the way in which the addition of noise affects the spread of disinformation. 

\begin{figure}[H]
    \centering
    \includegraphics[width=6cm]{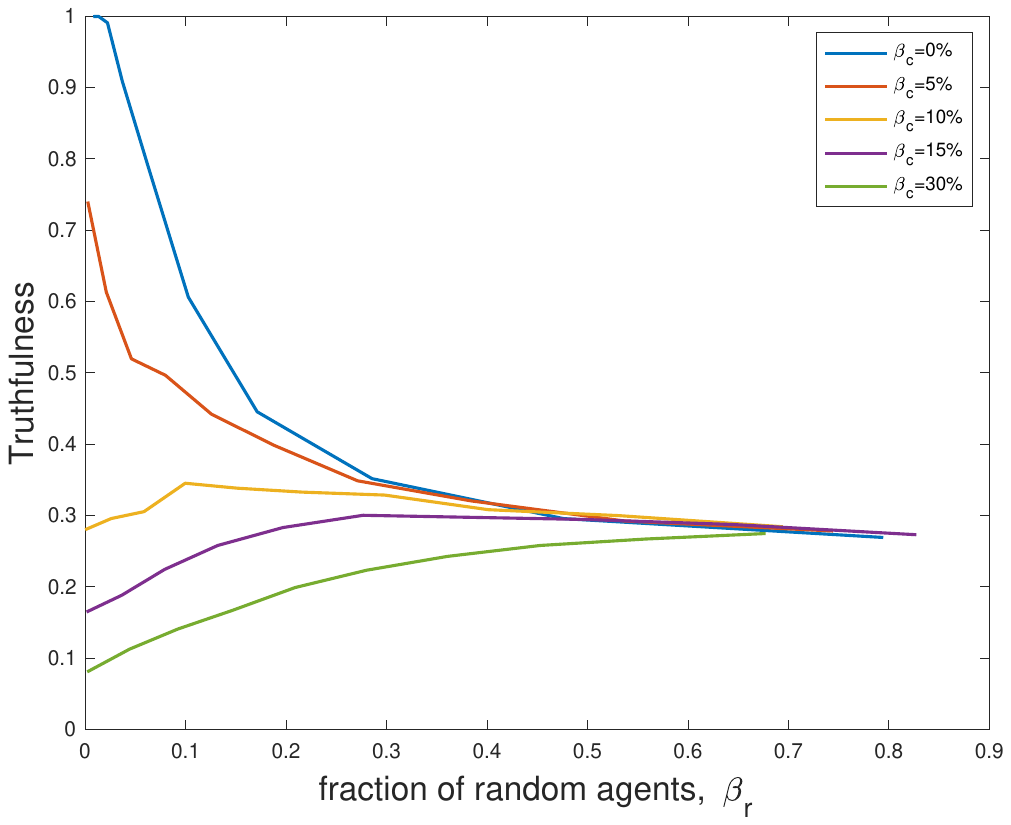}
    \includegraphics[width=6cm]{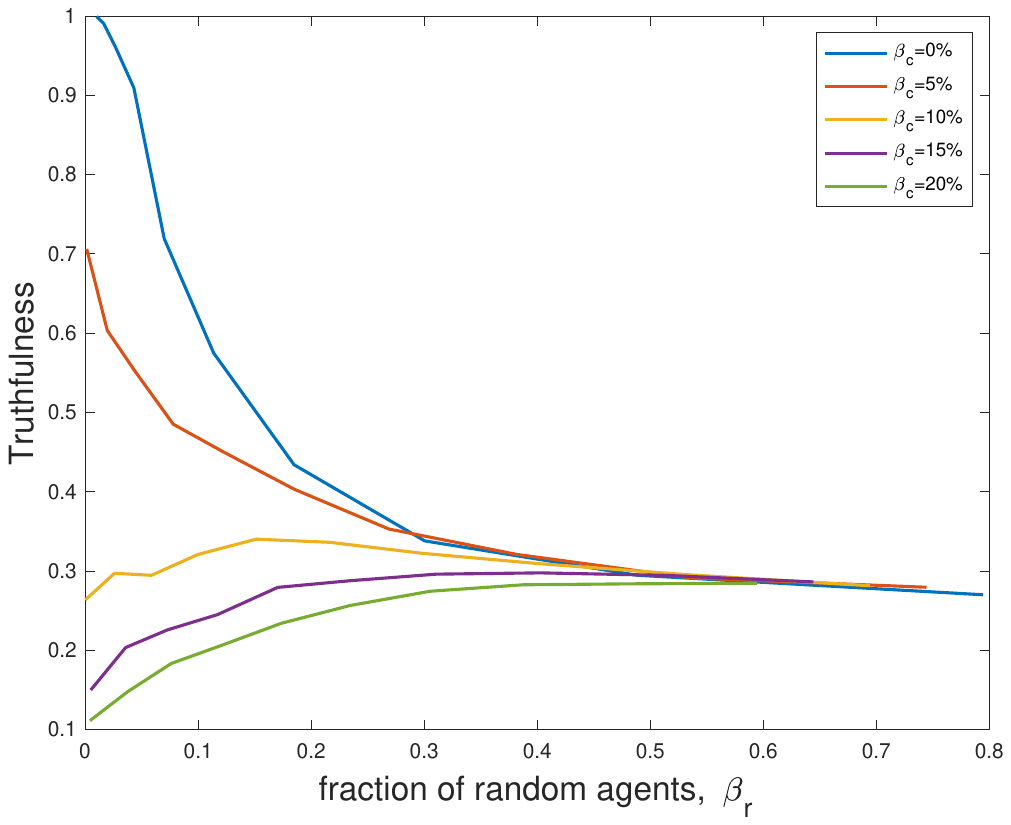}
    \caption{Average truthfulness as a function of concentration of agents who behave randomly with $N = 100$ agents and $M = 4$ hypotheses for an array of $\beta_c$ values, where noise is introduced through the Dirichlet policy (top panel) and the rewiring policy (bottom panel) for a series of ER networks with average degree of $k=10$.}
     \label{fig:Tr_1}
\end{figure}

In Figure~\ref{fig:Tr_1}, we observe truthfulness as a function of the fraction $\beta_r$ of random agents whose beliefs are generated via the Dirichlet (right panel) and rewiring (left panel) policies at different concentrations of conspirator agents, $\beta_c$. Depending on the presence of disinformation being pumped into the network, varying effects may be witnessed on truthfulness as a function of $\beta_r$. Specifically, both policies induce similarly positive and beneficial effects for larger values of $\beta_c$, i.e., we see that the injection of noise mitigates the effects of disinformation by increasing truthfulness. Intuitively, this may be seen to be due to a greater degree of `diversity’ of beliefs, essentially acting as a buffer against the disinformation presented by conspirator agents. Conversely, for lower values of $\beta_c$ we see an opposite effect, i.e., a reduction in truthfulness.

\begin{figure}[H]
    \centering
    \includegraphics[width=6cm]{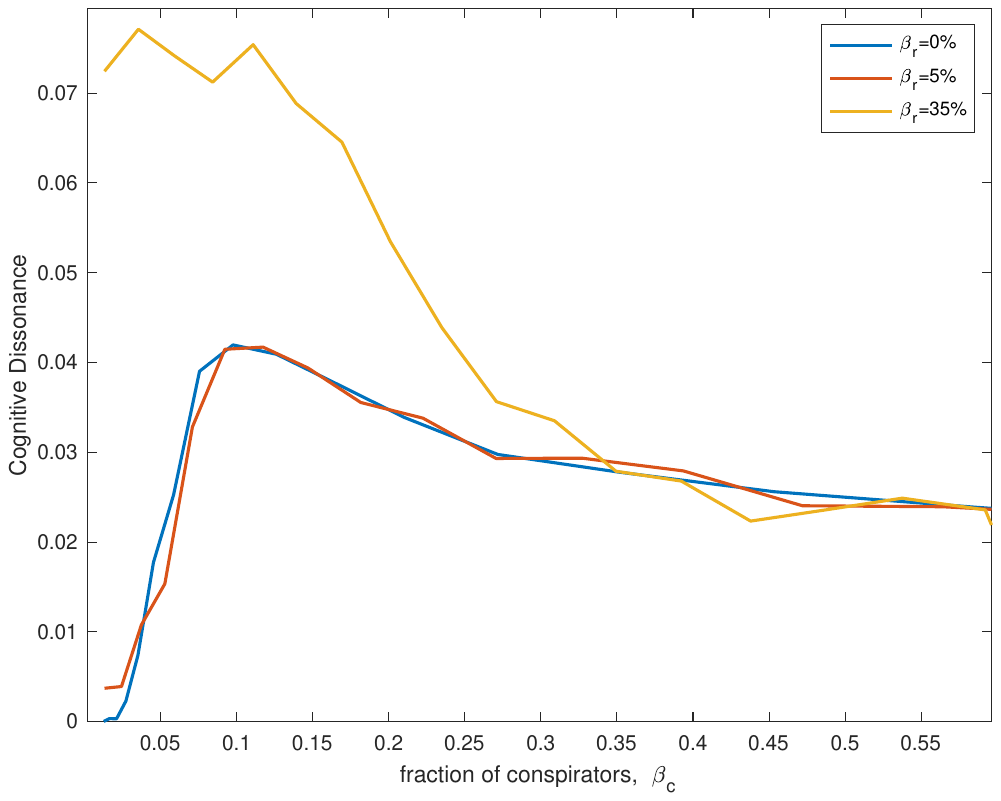}
    \includegraphics[width=6cm]{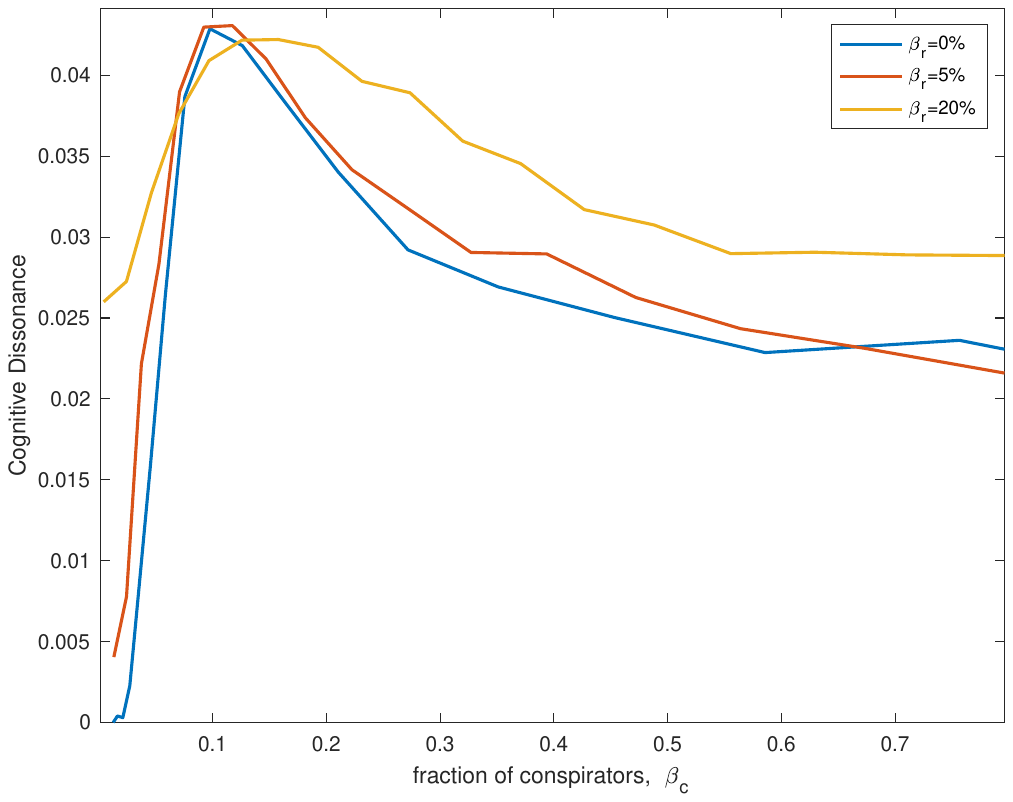}
    \caption{Average CD as a function of the concentration of conspirator agents $\beta_c$, for different values of $\beta_r$, where noise is introduced through the Dirichlet policy (top panel) and the rewiring policy (bottom panel) for a series of ER networks with average degree of $k=10$.}
     \label{fig:cd_1}
\end{figure}

Shifting our attention to CD, taken to be the difference between what a regular agent publicly expresses and privately believes, it was found in~\cite{riazi2024public} that agents in a social network typically experience the highest mental toll for fractions of conspirators in the network around $\beta_c=10-15\%$, and that CD subsides as the fraction of conspirators grows. As can be seen in Figure~\ref{fig:cd_1}, this non-monotonicity is essentially maintained under both policies, except for the combination of high $\beta_r$ / low $\beta_c$ under the Dirichlet policy, where we see considerably higher values of CD.

\begin{figure}[H]
    \centering
        
       \includegraphics[width=6cm]{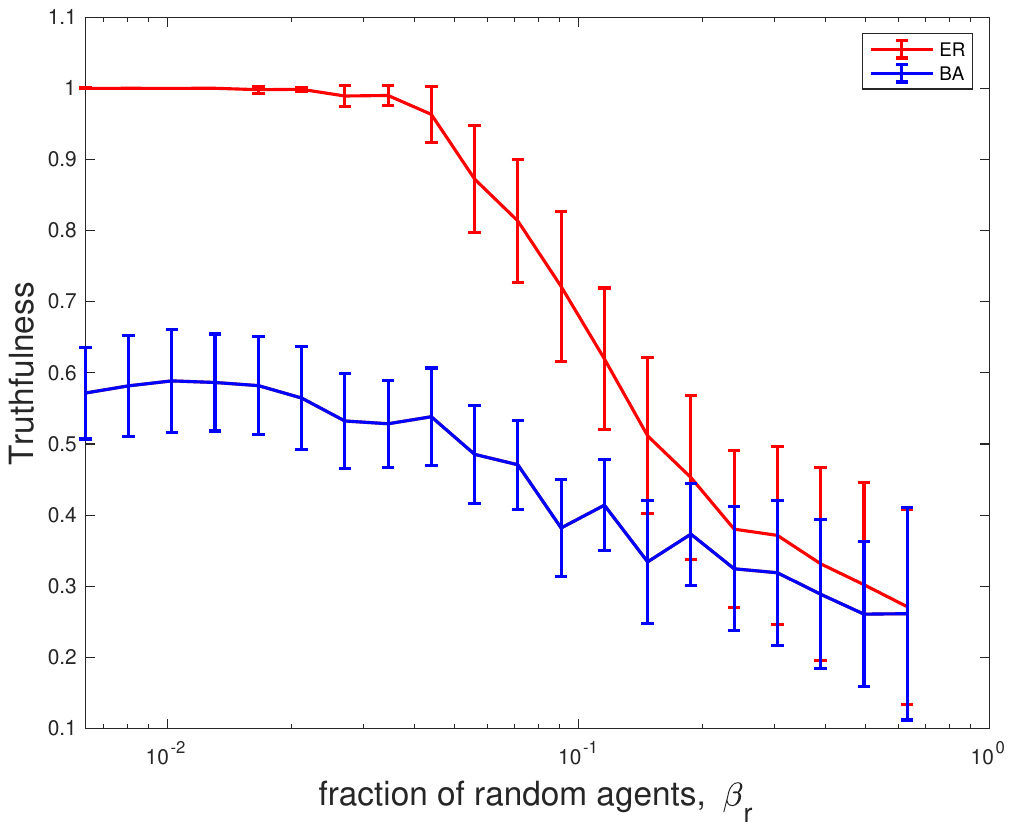}
        \includegraphics[width=6cm]{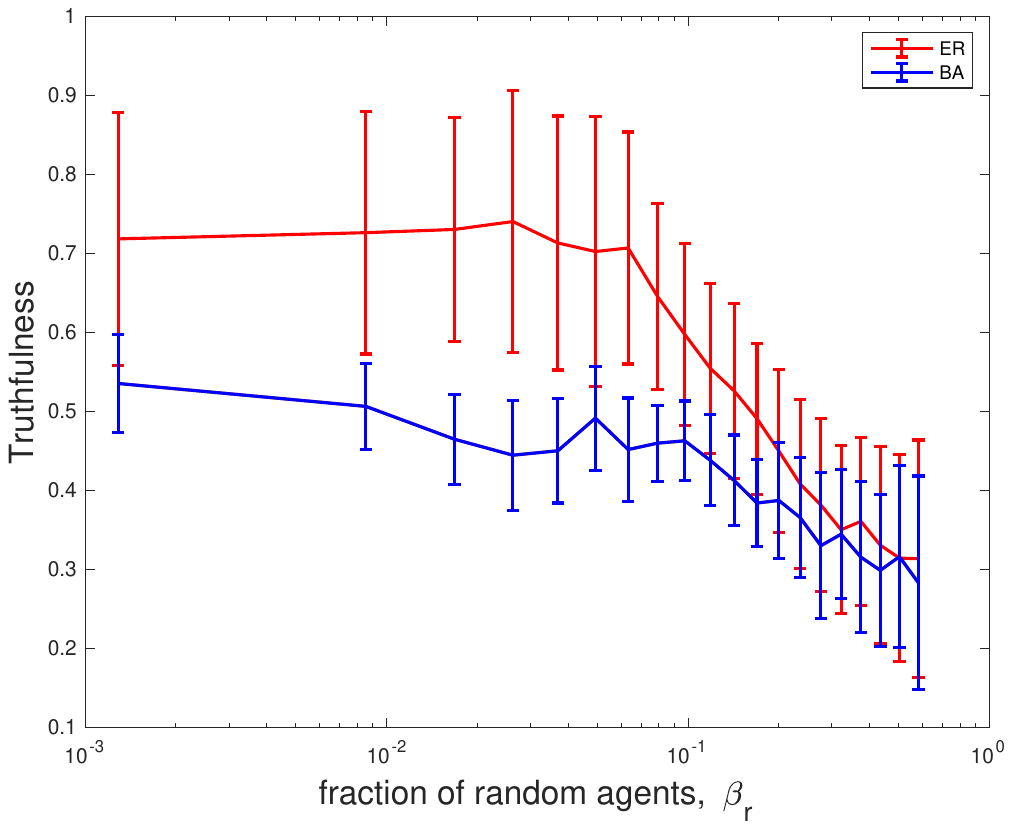}

    \caption{Average truthfulness as a function of the concentration of random agents $\beta_r$ generated via the Dirichlet policy for a series of BA and ER networks, with errorbars capturing simulation to simulation differences, where the left panel has $\beta_c = 0\%$, while the right panel has $\beta_c = 5\%$}
    \label{fig:3}
\end{figure}

\begin{figure}[H]
    \centering
    \includegraphics[width=6cm]{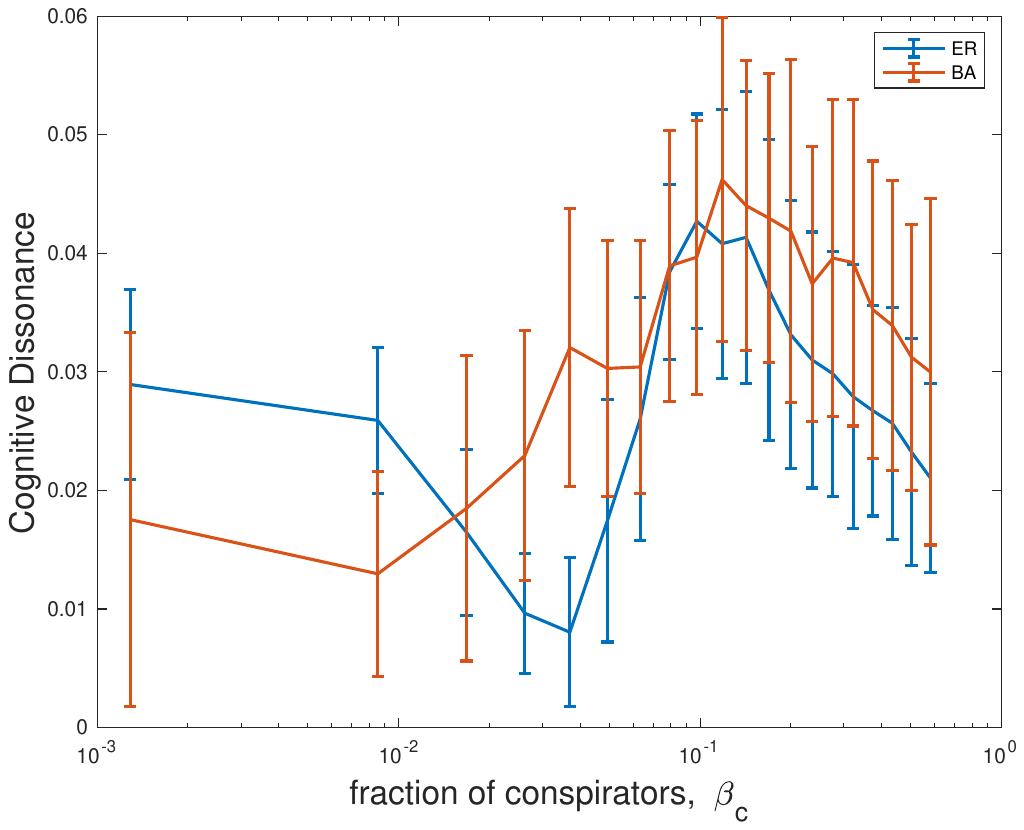}
    \caption{Comparison of CD as a function of concentration of conspirator agents at  $\beta_r=5 \% $  for a series of BA and ER networks, with $N = 100$ agents and $M = 4$ hypotheses, with noise generated via the Dirichlet policy.}
     \label{fig:4}
\end{figure}

In addition, we seek to understand which network topology is more resilient against the presence of disinformation. In order to do that, we compare results obtained in ER networks against those obtained in Barabasi-Albert (BA) networks~\cite{barabasi1999emergence}, which represent a well-known benchmark for networks with a heterogeneous, heavy-tailed, degree distribution. From Figure~\ref{fig:3}, it is apparent that truthfulness as a function of $\beta_r$ behaves similarly across the two network types. We perform comparisons in a scenario without conspirators ($\beta_c = 0\%$, left panel) and with conspirators ($\beta_c = 5\%$, right panel), implementing the Dirichlet policy. We notice that the two network types become effectively statistically compatible for larger values of $\beta_r$. However, for lower values of $\beta_r$ we see that the beneficial effects of randomness are perceived more strongly in ER networks. For the CD counter-part, seen in Figure~\ref{fig:4}, we see an effective statistical compatibility across the two network types, regardless of the specific values of $\beta_r$ and $\beta_c$.

\begin{figure}[H]
    \centering
    \includegraphics[width=6cm]{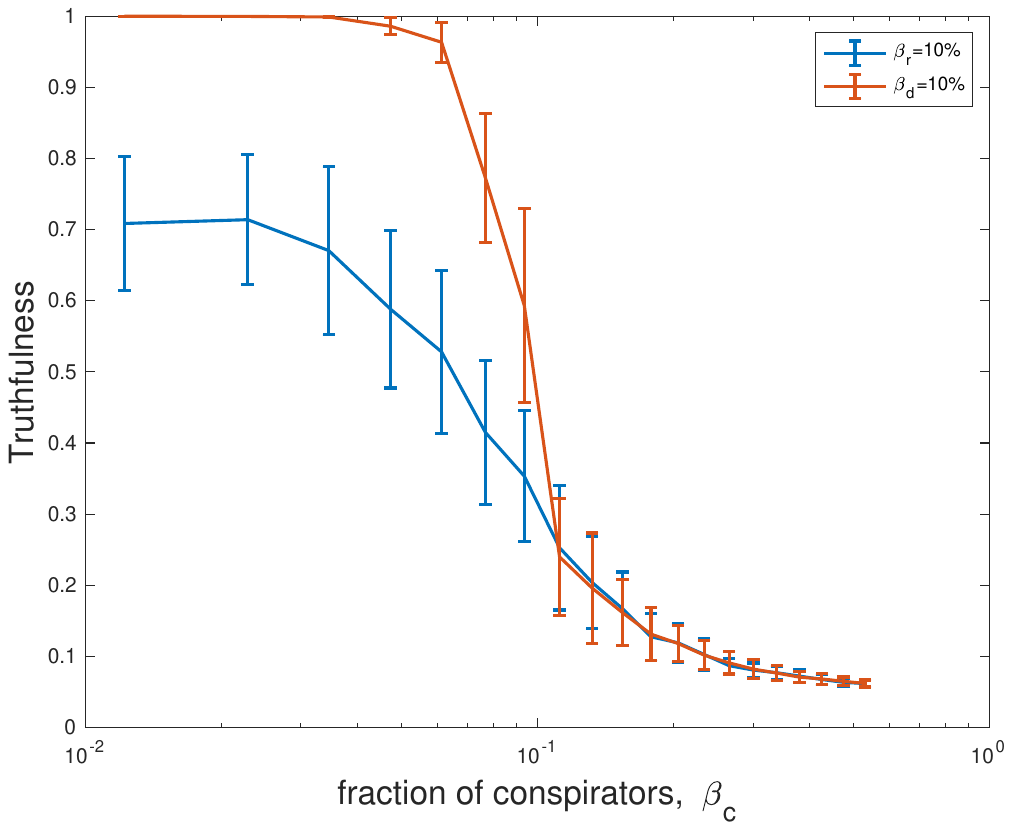}
    \includegraphics[width=6cm]{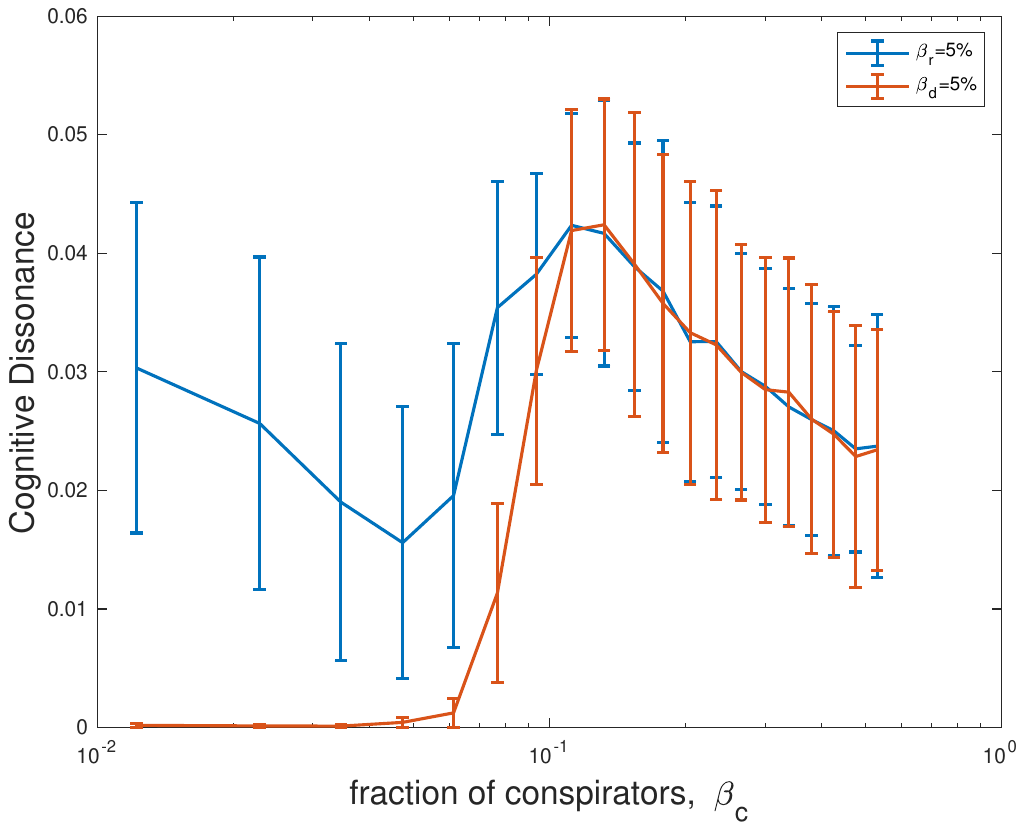}
    \caption{Truthfulness (left) and CD (right) as a function of the concentration of conspirator agents, comparing between sub-populations of $\beta_r=10\%$ noisy agents (via Dirichelet) and $\beta_d=10\%$ debunker agents,  with $N = 100$ agents and $M = 4$ hypotheses for a series of sparse ER networks $k=10$.   %where for a sub-population, beliefs are randomly generated via the Dirichlet distribution 
    }
     \label{fig:5}
\end{figure}

As mentioned, debunking, i.e., the act of exposing misleading information which may be in circulation within networks, is a common way of combating mis/dis-formation, as practiced by social media companies \footnote{ https://help.twitter.com/en/resources/addressing-misleading-info}. Thus, we now additionally seek to compare  the manner in which noise affects the propagation of disinformation with the effects which debunking generates. Following~\cite{riazi2024public}, we define a debunker agent as a node who actively promotes the ground truth, which we implement by imposing their private belief vector to be such that $b_M^{(t)} \approx 1$ and $b_j^{(t)} \ll 1$ for $j = 1,\ldots,M-1$ (i.e., debunker agents keep pushing the hypothesis corresponding with the ground truth with their publicly stated beliefs). In Figure~\ref{fig:5}, we observe that for a sufficiently high concentration of conspirators, injecting noise becomes as effective as debunking, both in terms of promoting truthfulness (left panel) and reducing CD (right panel).

\section{Conclusion}
In this paper, we sought to explore the role noise plays in combating disinformation in social networks. With spanning varying contexts of complex systems, where noise has been shown to be beneficial to build up resilience. Here we have made considerations such that to determine whether the same concept applies to social learning and diffusion of information. In a nutshell, our findings are in line with Taleb's point on antifragility~\cite{taleb2014antifragile}, i.e., that there is a tradeoff between efficienty and resilience in complex systems.

Invoking the DHT framework initially introduced in~\cite{lalitha2018social} and further explored in~\cite{riazi2024public}, we showed that depending on the concentration of conspirators present (and thus effectively the presence of disinformation), noise may be beneficial in boasting truthfulness, that is, the collective belief in the ground truth. However, such benefit may come at the price of a heightened cognitive dissonance, i.e., a more pronounced disconnect between their privately held and publicly stated beliefs, which we already demonstrated to appear in DHT models in~\cite{riazi2024public}.

We tested whether the above results are affected by network topology by running simulations in homogeneous (Erdos-Renyi) and heterogeneous (Barabasi-Albert) networks. On a qualitative level we found the overall behaviour to be similar across network types. However, on a quantitative level we found homogeneous networks to be more positively affected by the injection of noisy agents. This may be expected given that heterogeneous networks obviously feature hubs, which can enhance the propagation of disinformation. 

Finally, we compared the beneficial effects induced by noise to those that can be achieved via debunking, modeled as a fraction of agents actively pushing the ground truth. We found debunking to be more effective, but only for lower concentrations of conspirator agents (i.e., under mild disinformation). Conversely, we found the injection of noise to be as effective as active debunking for higher concentrations of conspirator agents. This is a rather remarkable point, as it is now well established that debunking has serious downsides, often referred to as the `backfire effect' \cite{redlawsk2002hot,nyhan2010corrections}. Our results suggest that, when disinformation is especially prevalent, injecting noise could achieve the same results as debunking, without its unintended consequences.

Most of our results are consistent across two different representations of noise, namely a sub-population of agents' beliefs being determined via the Dirichlet distribution as well as topological rewiring. In essence, there is no clear `winner' in terms of policy. 
Such configurations of noise may be analogized to different scenarios of arbitrariness, with the former corresponding to `bots' being introduced to a network, and the latter to algorithms with recommending content typically not interacted.

\bibliography{aipsamp.bib}% Produces the bibliography via BibTeX.

\end{document}